# State of the Art Review for Applying Computational Intelligence and Machine Learning Techniques to Portfolio Optimisation

**Evan Hurwitz**
School of Electrical and Information Engineering
University of the Johannesburg
Johannesburg, Gauteng, South Africa
hurwitze@gmail.com

**Tshilidzi Marwala**
Faculty of Engineering
University of Johannesburg
Johannesburg, Gauteng, South Africa
tmarwala@uj.ac.za

## Abstract

*Computational techniques have shown much promise in the field of Finance, owing to their ability to extract sense out of dauntingly complex systems. This paper reviews the most promising of these techniques, from traditional computational intelligence methods to their machine learning siblings, with particular view to their application in optimising the management of a portfolio of financial instruments. The current state of the art is assessed, and prospective further work is assessed and recommended.*

**Keywords:** reinforcement, learning, temporal, difference, neural, network, portfolio optimisation, genetic algorithm, genetic programming, markowitz portfolio theory, black-scholes, investment theory.

## 1 Introduction

This paper is to serve as a literature review for a Phd thesis to be carried out at the University of Johannesburg. This paper is to examine the state of the art of machine learning and computational intelligence techniques to the task of portfolio optimisation. First the background of what precisely what constitutes computational intelligence and machine learning techniques will be explored, in addition to the successful implementations and applications of these techniques, and the facets that make them so promising. Following on that, the problem of portfolio optimisation will be defined, including current objections to the standard model . Current applications and research will then be evaluated, after which limitations will be examined, both technical limitations and industry limitations. Finally, further research will be recommended based on the findings detailed in the paper.

## 2 Background

The field of portfolio optimisation is a field of immense importance to the economy of a country and indeed the global economy in general. Like it or not, the majority of stored money in the world, ranging from money in funds, in banks, in insurance schemes and more, is stored in portfolios of some sort or another, and therefore subject to all of the risks that financial markets pose. The management of these portfolios has therefore been a critical area of research ever since the inception of the marketplace, and the importance of this research is only increasing as financial instruments become ever more complex. The recent sub-prime crisis only further illustrates that our understanding of the systems we have created need further examination and understanding if we are to navigate these waters without bouncing from reef to reef.

The fields of computational intelligence and machine learning are concerned with the modelling of systems, and in particular with modelling systems that prove difficult, if not impossible to model by conventional mathematical means. In the case of computational intelligence, this is done by observing and optimising hueristics based upon gathered data. Conversely, the field of machine learning relies on learning from observation, oftentimes involving online learning, ie learning while the process to be modelled is in action / operation. Both of these disciplines require relatively intensive computing power, and as such are only coming into their prime now, owing to the advances that have been made in the past two decades in computing power, and in particular said computing power's ready availability to researchers.

# 3 Computational Intelligence

A number of computational intelligence techniques will be examined in this section, including their strengths, weaknesses and typical applications. The broad purpose of these techniques is to solve either classification or regression problems in order to model a given system. These models can be used for prediction purposes, or further than that for system control purposes.

## 3.1 Genetic Algorithms (GA)

Genetic algorithms attempt to imitate the evolutionary process of genetic evolution in order to optimise a given problem [1]. The problem is formalised by defining a *fitness function*, which determines a given value for the set of parameters to be optimised in such a way that maximising the value of this *fitness function* returns the optimal set of parameters [1]. The Algorithm derives its name from the manner in which it optimises its parameters. This is done by generating an initial *population* of possible solutions, and evaluating the fitness function for each possible solution [2]. These solutions are then used to generate a new population (each new population termed a *generation*), new individual solutions being generated by inheriting properties from its specific parents (choosing of parents can be done in many ways, as can the inheriting of features, but invariably the individuals with better fitness values are favoured by the chosen process) [2]. The precise choices of parent-selection-criteria and feature inheritance, as well as type of *mutation* (a random changing of a given value, used to explore further into the search-space) are all optimisation parameters that can be fine-tuned to optimise the process itself [2].

Genetic algorithms have found much success in optimising large searching systems that are difficult to quantify, and have in particular been very successful in the solving of Scheduling and Routing problems, in addition to financial applications [3].

A relatively recent offshoot of genetic algorithms is *genetic programming* [4], in which a problem-solving method, or policy, is evaluated and optimised by means of a genetic algorithm [4]. This in particular shows much promise in the Portfolio Optimisation realm for being able to combine multiple portfolio optimisation strategies into one coherent, optimised strategy.

## 3.2 Particle-Swarm Optimisation (PSO)

Particle-swarm optimisation is another method of solving the same types of problems as genetic algorithms, based on the paths of swarms rather than on the workings of genetics [5]. A *fitness function* is defined in the same way as with a GA, but the optimisation instead follows the following formulae for the velocity and position of each *particle* (solution) [5]:

$$v_{i,j} = c_0 v_i + c_1 r_1 (GB_j - x_{i,j}) + c_2 r_2 (LB_{i,j} - x_{i,j}) + c_3 r_3 (LB_{i,j} - x_{1,j}) \qquad (1)$$

$$x_{i,j} = x_{i,j} + v_{i,j} \qquad (2)$$

Where $C_i$ are learning constants, and $R_i$ being random vectors, typically between 0 and 1, and GB and LB represent Global Best and Local Best respectively [5].

Owing to its similarity to GAs in terms of function, PSOs have typically been applied to similar problems as GAs [6]. Both have also been applied to Portfolio Optimisation problems in various forms [7].

## 3.3 Neural Networks (NN)

The fundamental building-blocks of neural networks are *neurons* [8]. These neurons are simply a multiple-input, single-output mathematical function [8]. Each neuron has a number of *weights* connecting inputs from another layer to itself, which are then added together, possibly with a *bias*, the result of which is then passed into the neuron's *activation function* [8]. The activation function is a function that represents the way in which the neural network "thinks". Different activation functions lend themselves to different problem types, ranging from yes-or-no decisions to linear and nonlinear mathematical relationships. Each *layer* of a neural network is comprised of a finite number of neurons. A network may consist of any number of layers, and each layer may contain any number of neurons [8]. When a neural network is run, each neuron in each consecutive layer sums its inputs and multiplies each input by its respective weight, and then treats the weighted sum as an input to its activation function. The output will then be passed on as an input to the next layer, and so on until the final output layer is reached. Hence the input data is passed through a network of neurons in order to arrive at an output. Figure 1 illustrates an interconnected network, with 2 input neurons, three hidden layer neurons, and two output neurons. The hidden layer and output layer neurons can all have any of the possible activation functions. This type of neural network is referred to as a *multi-layer perceptron* [8], and while not the only configuration of neural network, it is the most widely used configuration for regression-type problems [8].

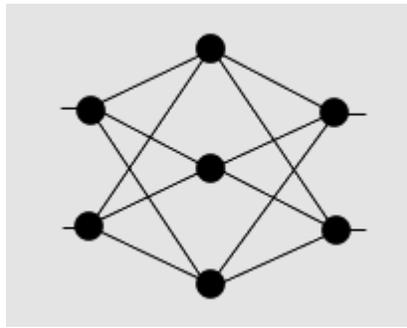

Figure 1. Sample Connectionist network

Neural networks have a number of characteristic properties that mark them as ideal for modelling. Specifically, these characteristics are [9]:

- Universal approximators

- Generalisation

- Pattern Recognition

Neural Networks have been applied in a multitude of differing topologies, and to a wide variety of problems [10]. Many attempts have been made to apply Neural Networks to the problem of predicting stock-market prices, although with very limited success [11]. One problem that is often overlooked when applying neural networks is the implicit assumptions that data structuring has on the underlying problem – for example, many NN stock-market attempts have assumed a simple time-series approach, making the implicit assumption that the historical prices of a single equity are sufficient to predict its future value [12].

## 4 Machine Learning

Machine learning is a subset of computational intelligence [13]. It is a field specifically focusing on allowing a system to maximise some reward signal, and thereby learn to operate within its environment [13]. The complexity of the modelling system can vary greatly, allowing the researcher a high degree of customisable freedom, with appropriate trade-offs inherent to each specific choice [13].

## 4.1 Reinforcement Learning

Reinforcement learning is a specific form of machine learning, in which the system can only react to the reward signal it is given, and is not told by the engineer how to react given a specific result [13]. A typical reinforcement learning problem will have the following topology:

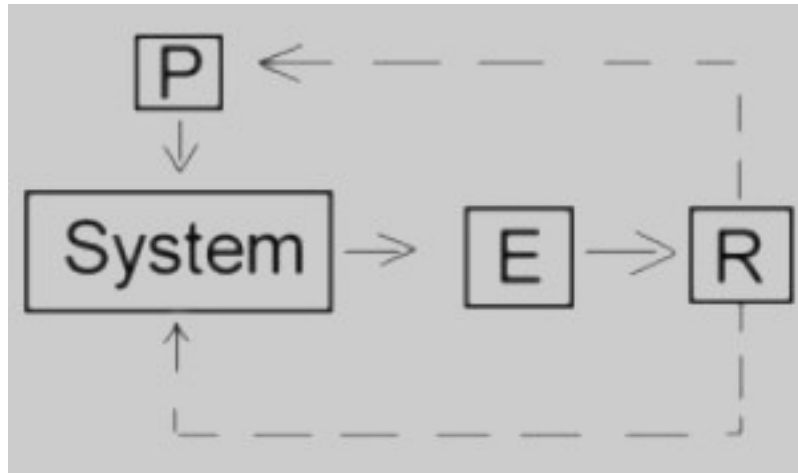

Figure 2. Reinforcement Learning Topology.

The reward signal R indicates the success / failure of the attempted policy's action within the environment, as well as the degree thereof. The learning algorithm L defines how either the policy parameters are updated based on the reward signal R, or the environment model E is updated with respect to R. In this way, the policy P learns to exploit the environment with which it interacts, in fact bypassing the act of modelling the system and instead directly modelling its own control system for working within the system [13].

## 4.2 On-Policy Versus Off-Policy Learning

Depending on the level of generalisation built into the model, a reinforcement learning system may learn either *on-policy* or *off-policy* – that is, the learning may be specific to the policy tested in order to acquire a given reward signal, or alternatively the learning may generalise to situations beyond the specific policy that has been used. The advantage of off-policy learning is that one can achieve far swifter learning, since every single scenario need not be encountered in order to learn to handle it [14]. The downside of off-policy learning is that learning can have unintended consequences since it applies to scenarios beyond that which has been observed [14]. In addition, models and learning algorithms both require greater complexity in order to successfully learn off-policy [14].

## 4.3 Tabular Learning / Dynamic Programming

This form of reinforcement learning is the simplest form. Each scenario is evaluated, and iteratively updated using an appropriate equation (such as the Bellman Equation) [14], in order to create a perfect model of the environment [14]. From this, a policy can be constructed to take advantage of the environment, picking the choice with the highest expected return.

## 4.4 Temporal Difference Learning

Reinforcement learning involves the training of an artificial intelligence system by means of trial-and-error, reflecting the same manner of learning that living beings exhibit [14]. Reinforcement learning is very well suited to episodic tasks [14]. In reference to *figure* 2, Policy P would be represented by some appropriate function approximator, which would then be updated according to the chosen learning algorithm [14]. This methodology allows for real-time learning, and also eliminates the need for expert knowledge [14].

Recent work by Sutton, Svepesvari and Maei [15] has proven an algorithm for off-policy learning with a linear function approximator using temporal difference learning and allows for linear complexity in memory and per-time-step computation [15]. Current learning algorithms still struggle to remain stable when dealing with nonlinear function approximators [16] [17].

## 5   Portfolio Theory

Portfolio theory deals with the problem of allocating funds within a portfolio [18]. This problem is generalised to include any defined universe of financial instruments [18]. The goal of the portfolio optimisation problem is to achieve maximum return on investment for the minimum amount of risk [18]. Optimising this risk-reward trade-off forms the heart of portfolio theory, which is quite appropriate as this is the key question in the portfolio management industry, namely managing clients' funds in such a way as to keep an appropriate level or risk while maximising the clients' return on investment.

### 5.1   Modern Portfolio Theory (MPT)

Centered around Markowitz' theories [18], this forms the central tenet of modern investment theory [18]. Price moves within the market are assumed to be independent, stochastic events [18] and are then treated as *random-walk* problems [18] (in some literature referred to instead as *Markov Decision Processes*) following typical *Brownian Motion*. With this model, portfolios are then examined through a statistical framework, modelled only by two parameters, namely their *mean return* and their *standard deviation*, the latter of which is interpreted as the inherent riskiness of the examined asset [18]. By quantifying assets according to these two parameters, assets and strategies can be compared on a risk-adjusted reward basis, with *efficient* strategies being the highest reward for a given amount of risk [18]. Alternatively, fund managers of a targeted-return fund will attempt to achieve a specific return, striving for the lowest possible risk in a given portfolio. These optimal (*efficient*) strategies lie along the *efficient horizon* [18] as indicated in Figure 3.

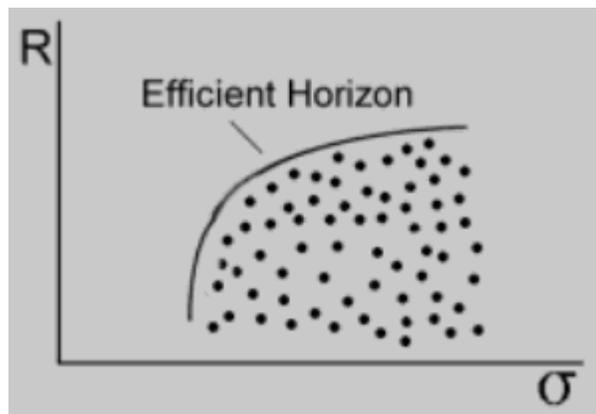

Figure 3. The Efficient Horizon, with strategies plotted according to return (R) and risk (σ)

Unfortunately, the assumptions within the Markowitz framework have been seen to be invalid within the investment arena [19] [20] [21]. In particular, assumptions of independence, both to internal historical moves, and to external moves, have been seen to be false [21]. Similarly, the assumption of Brownian Motion in price moves does not hold up to scrutiny and in most cases grossly underestimates the extent of possible price swings [21], with the industry standard approach being to discard such events as outliers [21], even though they occur with too high a frequency to be validly treated as such [21].

### 5.2   Post-Modern Portfolio Theory

If one accepts the premise, as do many investors and investment houses, that returns above the expected return are not risky (or an indication of risk in the underlying instrument) then the use of variance as a measure of risk is in fact inappropriate [22]. Markowitz in fact recommended the use of *semivariance* as a measure of risk, but was of the opinion that the computational difficulties were too formidable [24]. This, unsurprisingly, is certainly no longer the case. Semivariance is a measure of the downside variance of a dataset, excluding values that are above the mean from the risk calculation [22]. With a different measure of risk formulated, one can then apply MPT as before, but now designed to only treat losses as risks, not gains [23].

### 5.3 Quantifying Risk

The task of quantifying risk is one of extreme importance to the financial industry, yet also extremely difficult [25]. In the field of portfolio management risk is specific to the amount of money one can expect to lose should the market not act favourably [25]. The traditional measure of risk, as has been mentioned in previous sections, is the statistically measurable standard deviation of a portfolio [22]. Application of this risk measure, however, uses the assumption of a normally distributed price moves [21], which has been shown historically to be an inaccurate and insufficient method of measuring risk [21]. One suggested method, recommended by Mandelbrot, is to use a *fractal index* as a measurement of risk, which is conceivably more reasonable due to it possessing a scaling property more indicative of real price moves than the static Gaussian distribution [21]. Note that this only applies to *volatility* risk, which does not encompass the entirety of the risk involved in investing [21].

## 6 Portfolio Optimisation

Portfolio optimisation is typically done by choosing a specific strategy (i.e. a mid cap fund, targeted return, etc) and then optimising the risk-reward ratio within the limitations of the particular strategy [26]. The strategy chosen will constrain the available instruments for the portfolio, while thereafter the optimisation will select from those available to create a portfolio [26]. As an example, consider a fund investing money on behalf of a pension fund. The money is required to grow, but is legally prohibited from *selling short*, and is mandated by the clients to have a relatively low level of risk, on the grounds that preserving capital is more important than superior growth. Lastly, if we assume the fund is a large fund, it will also have a liquidity limitation, on the grounds that if the fund manager should decide an asset needs to be sold, that buyers can be found for the amounts being held. (Small penny stocks are unlikely to be traded in the quantities required, and hence would be disqualified from the fund's *universe of stocks*). The Manager would then plot the remaining instruments and strategies on a graph of risk versus reward (as in Figure 3) and then compile a portfolio based upon his research in order to comply with his given mandate [26]. This process will need to be re-done on a regular basis as the various instruments' performance changes (changing their ratings) and as profits and losses change the composition of the portfolio, affecting its overall risk-reward ratio.

## 7 Applied Combinations

Owing to the promising properties of machine learning and computational intelligence techniques, numerous attempts have been made to apply these techniques in the investment arena, with widely varying rates of success. Predictive methods have been attempted in numerous different formats, using genetic algorithms [27], neural networks [12] [28] and others. Attempts have also been made to utilise these techniques for option pricing [29], as an alternative to the Black-Scholes and Monte-Carlo modelling approaches that are the current industry norm. Recently, Rosen and Saunders [30] have examined using analytical techniques for pricing collaterolised debt obligations (CDOs), although they also fall into the old trap assuming a normally distributed data set (in this case the creditworthiness of the constituents of the portfolio).

The need for utilising systems that include the dependant nature of the market that the initial assumptions of Markowitz denies has been brought up again by Comte [31], looking specifically at long-term memory in volatility models.

Neural network applications within finance have even been attempted as far afield as in marketing campaigns, such as the work done in [33] where notably a neural network-based approach was used to assign discounts in a marketing campaign in order to maximise its effectiveness. Notably, the researchers utilised machine learning to optimise the policy, but still kept the neural network model static once it had been trained [32].

Neely [33] used a genetic programming approach to find optimum trading rules, working with a technical analaysis (TA) basis, but found that their results were inferior to a simple buy-and-hold strategy. Allen and Karjalienen attempted a similar approach [34], using a genetic programming to tune technical trading rules, but in this case found that their approach actually lost money when backtested. It is notable that both approaches utilise only historical data of the actual to-be-traded security as inputs, and thus adhere to the assumption that individual price moves are independent events (an assumption quite vigorously challenged, as indicated earlier). Lin et al [37] attempted a similar approach, using genetic algorithms to fine-tune the parameters in TA-based trading strategies and Filter Rules. This research was evaluated using simulated trading in the Australian Stock Market [37]. The method was compared with a *greedy* algorithm providing optimal returns for efficiency comparison. The GA proved to be far more efficient than the greedy algorithm, accomplishing in 10 seconds what the greedy algorithm achieved in 10 hours [37]. The testing appears to also have been on a limited selection of stocks, and thus somewhat suspect. Cura [38] in turn investigated using a particle swarm optimisation approach, and compared it to GA, tabular search and simulated annealing aproaches, using the Markowitz-mean model. While the PSO results were promising, no testing was done on real-world data [38].

More recently, Freitas et al [35] investigated using *auto-regressive moving reference neural networks* (AR-MRNNs) in order to predict price moves for short-term trading. They worked within the Markowitz framework, aiming to create an efficiently diversified portfolio, trading using the predictor based on the AR-MRNN. Not only did they manage to beat the mean-variance model, but also in fact beat the market (something that traditional portfolio theory states is impossible, based on the challenged assumptions detailed earlier) [35]. It is worth noting that even this highly promising result still used a relatively simple input-output structure, which limits the ability of the network.

The necessity of adapting to changing market condition was explored by Nagayama and Yoshii [36] as they explored forgetting appropriate data in a binary classifier, in order adapt the classifier to changing conditions by forgetting data that was no longer relevant. The explored both active and passive agent forgetting in their trials, with promising results. It should be noted, however, that the test data used is very limited, and the results could very well be a feature of the data.

## 8  Application Limitations

In the field of machine learning, it is still remarkably difficult to implement reinforcement learning when using nonlinear function approximators, such as neural networks [16]. This limitation makes application to nonstationary systems (as financial markets have shown themselves to be) a challenging endeavour. Sutton has recently managed to develop an algorithm for off-policy learning using temporal difference learning with a linear function approximator that allows for linear complexity in memory and per-time-step computation, almost but not quite enough to handle the nonlinear case of neural networks. A further limitation that has become apparent is the inadequacy of traditional risk measurement benchmarks, as their unrealistic assumptions taint the underlying analysis [21] [39]. The assumptions implicit in the Markowitz model are in fact no longer convenient, but now actually a hindrance to meaningful work. Much of the research is also out of touch with the realities of the investment world, concentrating on arbitrary measures of reward and only paying superficial attention to the issue of risk, which in reality dominates the

## 9  Recommended Research

In this field, there is much scope for meaningful work that, if current tumbling financial markets are any indication, is desperately needed. The potential for machine-learning and computational intelligence techniques to perform prediction-based and classification-based portfolio optimisation is readily apparent owing to their various properties, and the various niches are beckoning. The development of an algorithm to perform temporal difference updates that remains stable for nonlinear function approximators would be an invaluable advance. For the tasks of predicting; modelling and classifying the markets, the space is open for research with far greater input space, in turn making fewer inherent assumptions about the market that are unlikely to hold up to scrutiny. In particular, the astonishingly common assumption that historical prices / price moves of a given financial instrument or index is sufficient to model its behaviour is ludicrous – the assumption that all information necessary is contained within is a rigid interpretation of the efficient market hypothesis, and one that does not bear up to scrutiny. While all information may be reflected in a price, it is obfuscated by overlying effects, and does not account at all for longer-term dynamics in the system, which cannot be ignored if a meaningful / useful model is to be developed. Further research into risk measurement is required, with fractal indexing being a particularly promising field currently. Modifying traditional risk measures to make fewer assumptions about the market also promises to be a fruitful area of research. Lastly, using computational intelligence and machine learning approaches, an adaptive portfolio optimiser is theoretically feasible, combining multiple trading strategies to create and maintain an optimised portfolio according to modern portfolio theory measures, while utilising objective measures of risk without unrealistic assumptions.